# Model Validation Study for Central American Regional Electrical Interconnected System

Xiaoyuan Fan[1], Marcelo A. Elizondo[1], Pavel V. Etingov[1], Mallikarjuna R. Vallem[1], Shuchismita Biswas[1], Seemita Pal[1], Carlos Erroa[2], Christian Munoz[2], Daniel Polanco[2], Victor Villeda[2]

1. Energy and Environment Directorate
   Pacific Northwest National Laboratory
   Richland, WA
   Email: xiaoyuan.fan@pnnl.gov

2. Planning and Operation
   ENTE OPERADOR REGIONAL
   San Salvador, El Salvador
   Email: cmunoz@enteoperador.org

*Abstract*—The Central American Regional Interconnected Power System (SER) connects six countries: Guatemala, El Salvador, Honduras, Nicaragua, Costa Rica, and Panama, it is operated by the regional system operator *Ente Operador Regional* (EOR). Due to its geographical shape and layout of major transmission lines, SER has a weakly meshed grid, where disturbances can easily propagate, challenging its reliability. Having an accurate dynamic model is important for EOR when facing those reliability challenges. This paper describes interconnection-level model validation efforts for the SER and Mexico interconnected system. A detailed equivalent model of Mexico is incorporated in the existing SER planning model used by EOR. The resultant model is then validated using simulated dynamic contingency analysis and real system disturbance data. A fully automated suite of scripts is also developed and shared with EOR engineers. This work helps EOR improve their validation routine practices, to continuously improve SER dynamic model, and hence its reliability.

*Index Terms*--Power systems model validation, Synchrophasor data, Dynamic simulation, Remedial Action Scheme

## I. Introduction

Power systems models are used in numerous planning and operation studies. They need to be validated frequently to check how well they reflect existing operating conditions. Although validation of components like generators, governors and controllers is quite widespread, large interconnected system models are not commonly validated today. In recent years, deployment of Phasor Measurement Units (PMU) has provided utilities access to precisely time-stamped and synchronized measurements across large areas, presenting opportunities for validating existing models [1][2].

However, several challenges persist, including uncertainties in determining pre-event operating conditions, corruption of PMU data, and so on. Also, generating a precise sequence of events by analyzing field measurements is a demanding task. Without a fully automated procedure, it requires a significant amount of time and effort for utility companies to perform an interconnection level model validation. The Central American Regional Interconnected

This work was funded by the U.S. Department of State, Bureau of Energy Resources, Power Sector Program (PSP). PSP provides technical and regulatory support to the Central American Regional Electricity Market.

Power System (SER, *Sistema Electrico Regional*) connects six countries: Guatemala, El Salvador, Honduras, Nicaragua, Costa Rica, and Panama, and *Ente Operador Regional* (EOR) is the regional system operator. A schematic representation of the interconnection-wide layout is shown in Figure 1.

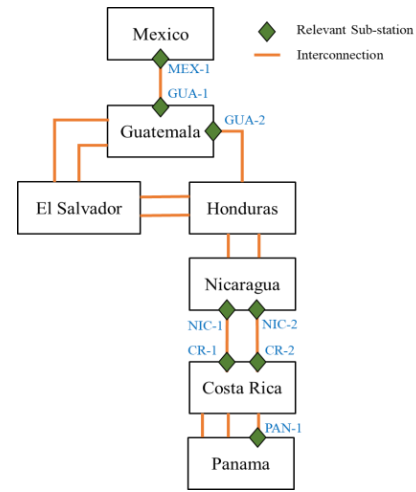

Figure 1. Layout of Central American Regional Electrical Interconnected System and its major transmission lines.

Reliable operation of the SER is coordinated between EOR and the national system operators and market operators (OS/OM, *Operadores de Sistema / Operadores de Mercado*), from the six SER member countries. According to the rules of the regional electricity market, when it comes to reliable operation, EOR is responsible for:

- Managing and coordinating the technical operation of SER with the national OS/OM
- Monitoring SER reliability by compliance of the security and performance criteria (CCSD, *Criterios de Calidad, Seguridad y Desempeño*)
- Determining reliability-related control actions through coordination and delegation of execution of these control actions to the OS/OMs. EOR is also responsible for determining necessary corrective actions in order to maintain the CCSD. Corrective actions may include load shedding, generation disconnection, and opening of transmission line
- Coordinating the investigation and analysis of reliability-related events in coordination with the OS/OM. EOR requests information related to these events from OS/OM and agents
- Calculating the operational transmission limits of regional transmission network (RTR, *Red de Transmisión Regional*), at least once a year using information from OS/OM and agents
- Validating or re-calculating the operational transmission capacity limits of the RTR and establishing the conditions and operational criteria necessary to maintain the CCSD
- Maintaining and gathering the necessary data for regional planning studies

Due to the geography of the region and layout of major transmission lines, it has a weakly meshed grid (shown in Figure 1) where disturbances can propagate quickly, posing reliability challenges. SER also faces difficulties in modeling interarea oscillations [3]. Mexico is connected to SER via a single point of interconnection (MEX-1 - GUA-1) at its border with Guatemala. Due to Mexico's relatively large size, this interconnection plays an important role in SER's stability. At present, EOR uses a Single Machine Two Load (SMTL) equivalent representation of Mexico in its planning model. However, this approach could not capture system behaviors with adequate accuracy, especially at the point of interconnection of two regional models; this deteriorates the situational awareness of grid operators as well as the design, review, and implementation of potential control schemes, such as protection relays setting and remedial action schemes (RAS).

In this work, the voltage and frequency behavior of the Central America and Mexico interconnected model is analyzed through dynamic contingency analysis, and further validated using real power system events. Major contributions of this work are:

- A detailed introduction of wide-area regional planning and event analysis process in Central America is provided.

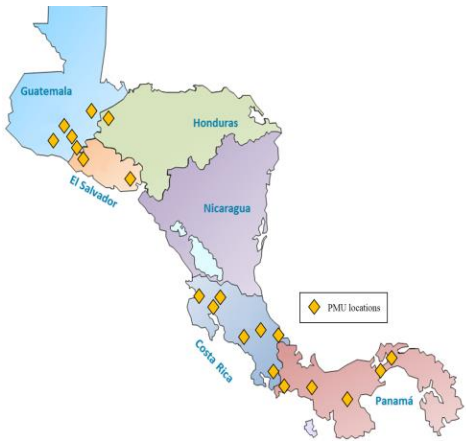

Figure 2. PMU locations in Central America by 2017.

- A suite of fully automated scripts is developed for the interconnected Central America and Mexico system model validation, including regional grid models integration, dynamic contingency analysis with flexible definition, dynamic model evaluation and PMU/SCADA data interface as well as visualization

The rest of the paper is organized as follows. Section I discusses prior instances of system wide model validations and associated challenges. Section II describes the modelling and characteristics of the Central American power grid. Section III elaborates the methodology used in this work. In Section IV performance of the interconnected Central America and Mexico model is demonstrated using a system disturbance event. Section V concludes the paper.

II. DESCRIPTION OF THE INTERCONNECTED CENTRAL AMERICA AND MEXICO SYSTEM

Despite being the next logical step after component level validations, wide-area models are not routinely validated at present. This is primarily due to the difficulty in acquiring time-synchronized data throughout a large system. With the relatively recent deployment of PMU technology, this challenge has been overcome and several entities have used PMU data following a disturbance to check the performance of system models [1] [2]. A qualitative scale for evaluating system models is described in [4]. In recent years, the SER member countries have installed many PMUs, and the recorded synchrophasor data has been utilized for different applications including inter-area oscillation analysis. Figure 2 provides the geographical locations of PMUs installed in six countries in Central America.

Reproducing a large power systems event, however, is not a trivial task. Constructing a precise sequence of events analyzing PMU and SCADA data is a herculean exercise, and is further compounded by limited information and/or potential measurement errors. Accurately determining pre-event operating conditions becomes even more difficult when coordination among multiple organizations is required. Any mismatch in the pre-event operating point can severely impact simulation results and hinder the identification of model inadequacies [5].

Considering the above challenges, it is apparent that rigid quantitative metrics cannot be used for evaluating a large power system model. Depending on the configuration and topology of a power grid, engineers might also choose to evaluate whether the system model faithfully describes a critical element (for instance, a major tie-line). The validation approach used by the authors is discussed in detail in Section III.

*A. Modeling Central America and Mexico Interconnection*

The Central America and Mexico interconnection connects the 400-kV system in Mexico with the 230kV system of SER. Due to Mexico's relatively large size, the level of transfer on this line is comparable to or higher than the largest generation contingency in SER as well as the capacity of lines interconnecting pairs of member countries. Therefore, losing this line can severely impact SER's operations [6].

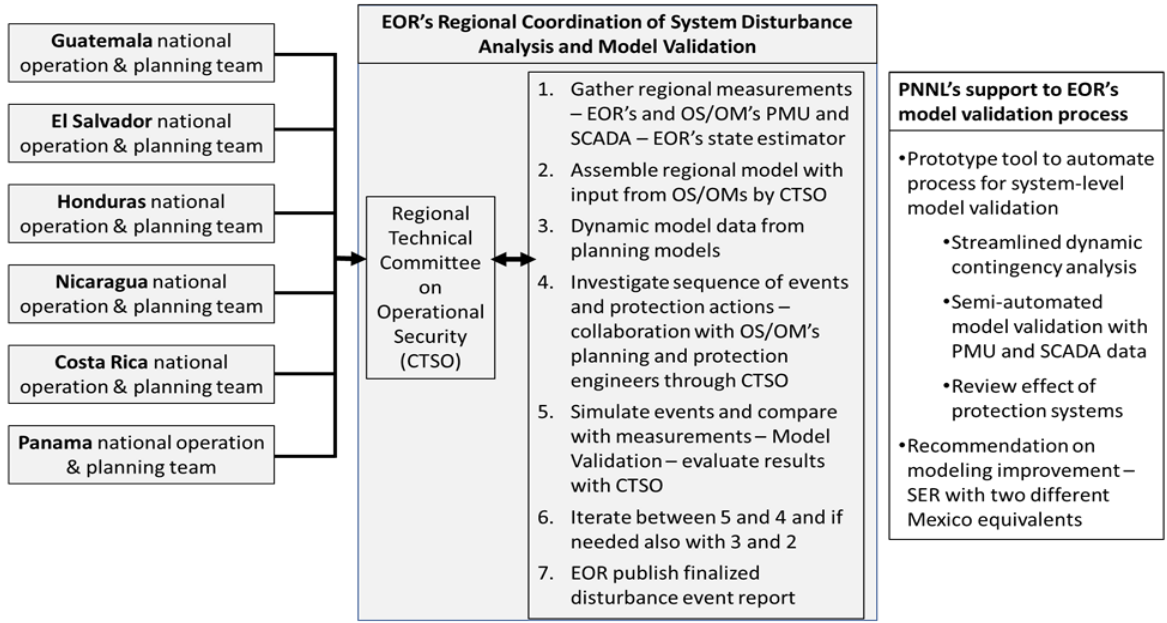

Figure 3. Relationship of this study to national transmission operation & planning collaboration on system disturbance analysis & model validation

*B. Modeling the Central American Regional Interconnected Power System (SER)*

The SER has experienced severe system outages producing blackouts in the region. EOR and the six national operators (OS/OM) have engaged in collaborative work to investigate the causes of the large outages and find mitigation options. The outages involved potential mis-operation of RAS (known in the region as "*sistemas de control suplementario*") and line distance relays.

Therefore, EOR engaged three working groups: Working Groups on Regional Coordination and Planning, Regional Coordination of System Disturbance Analysis, and the Regional Coordination of Operation and Protection.

A salient aspect of this collaboration is that it involved planning engineers that usually perform power flow and transient stability analysis from EOR and the OS/OM, as well as protection engineers from the OS/OM. These two types of engineers work with separate models and separate software solving problems that require unique skills, different from each other. But the nature of the outages in Central America needed this collaboration, which was achieved in the region, with EOR's leadership. Figure 3 illustrates this collaboration.

An important aspect of this collaboration is to build a model dataset that represents the pre-disturbance condition of SER. For this purpose, EOR requested each OS/OM to build a power flow model of their country with the information from their national SCADA system. Each OS/OM provided the models to EOR, and EOR assembled the full SER model.

EOR also collaborated with protections engineers from OS/OMs, to improve representation of outage sequences, such as tripping actions from RAS and protective relays like line distance relays.

*C. Remedial Action Schemes (RAS)*

i. *Mexico-Guatemala RAS:* The Mexico-Guatemala interconnection is protected by two relays- a) an oscillation relay acts if undamped oscillations persist for several seconds; b) a transfer-trip relay acts if two conditions are met- Mexico-Guatemala power transfer increases beyond a preset threshold and voltage at MEX-1 falls below a predetermined value. A large generation contingency in SER can result in the injection from Mexico to grow very rapidly.

ii. *Nicaragua-Costa Rica RAS*: This RAS protects an arterial transmission line in Nicaragua. For the purpose of this paper, this line is referred to as Line *N1*. If line *N1* is overloaded, a predetermined value of generation is first disconnected from the Nicaraguan region. If the overload continues to persist, then one of the tie-lines connecting Nicaragua to Costa Rica (NIC-1 - CR-1) will be tripped. Nicaragua and Costa Rica remain connected via the second tie-line NIC-2 - CR-2, even though the second one might also be subjected to the risk of overloading and tripping of protection relays.

The region has already undertaken efforts to evaluate the performance of existing RAS, as well as the possibility of implementing schemes at inter-area interfaces to arrest cascading events.

III. METHODOLOGY

In this section, the methodology adopted for different steps in the validation process are discussed in detail.

### A. Integrating Regional Models

At present, EOR uses a Single Machine Two Load (SMTL) equivalent model connected at GUA-1 to represent Mexico in the planning model of SER. In the simplified representation, all the generation and load in Mexico is concentrated at MEX-1. This approximation may produce distorted results in power system studies, mainly because the MEX-1 bus appears to have much stronger reactive power support than reality. The Central America only model with SMTL equivalent is also unable to replicate interarea oscillations. Considering the abovementioned difficulties, a previous study recommended incorporating a more detailed representation of Mexico in SER's model [7]. In this work, a better dynamic equivalent model of Mexico is integrated into SER's model, and additional revision on the models has been implemented, adding an interconnection branch and a two-winding transformer. A second parallel transformer is added at the Mexico-Guatemala interconnection to reflect recent infrastructure developments.

A fully automated suite of Python scripts was developed to accommodate those changes, as a result, now it is very easy to EOR engineers to use different SER planning/operational power flow cases for Central America and Mexico model integration. Figure 5 shows the comparison of voltage behavior at bus MEX-1 in one dynamic contingency analysis, which clearly shows better dynamic response from the improved model.

### B. Dynamic Contingency Analysis for CA+Mexico model

Validation is conducted in two stages. First, dynamic contingency analysis results obtained from the integrated model are compared with results from the CA-only model. There are in total 17 dynamic contingency analyses that were performed for different Mexico equivalent models, including 4 in Mexico, 10 in Guatemala, 2 in Panama and 1 in El Salvador. The dynamic contingency types include generator tripping, 3-phase to ground faults and line tripping. Compared with the Mexico SMTL Equivalent model, Figure 4 clearly shows better voltage and frequency behavior with the Mexico Improved Equivalent model.

### C. Model Validation Procedure for Real Disturbance Event

To assist Central American system-level model validation, PNNL has designed and implemented an automated process in Python through application programming interface (API) with PSS/E. The specific features of this automated procedure are given as follows:

1) Automated model integration of Mexico equivalent model and Central America grid model

2) Automated extraction and comparison of power flow conditions/measurements/quantities between SCADA/PMU data and simulated data for pre-event, during event and post-event stages

3) Automated governor parameter adjustment for flexible spinning reserve requirement

4) Automated power system event modeling based on system-level dynamic simulations

5) Automated PMU data processing and extraction [8]

6) One comprehensive procedure to identify potential mismatch in power flow conditions and erroneous model parameters in power system dynamic models

A schematic representation of the validation procedure is shown in Figure 6.

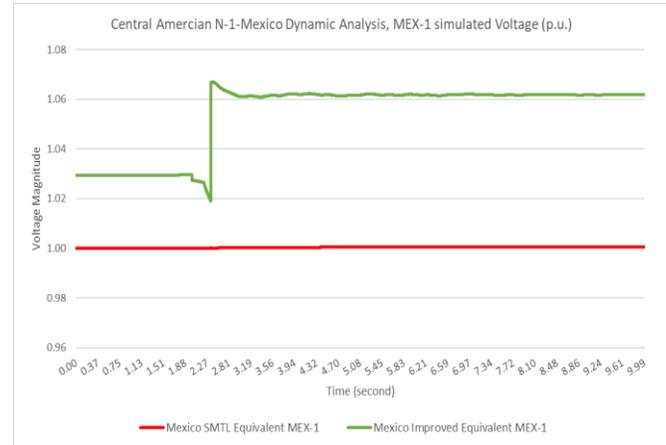

Figure 4. Voltage comparison of bus MEX-1 for different Mexico equivalent models.

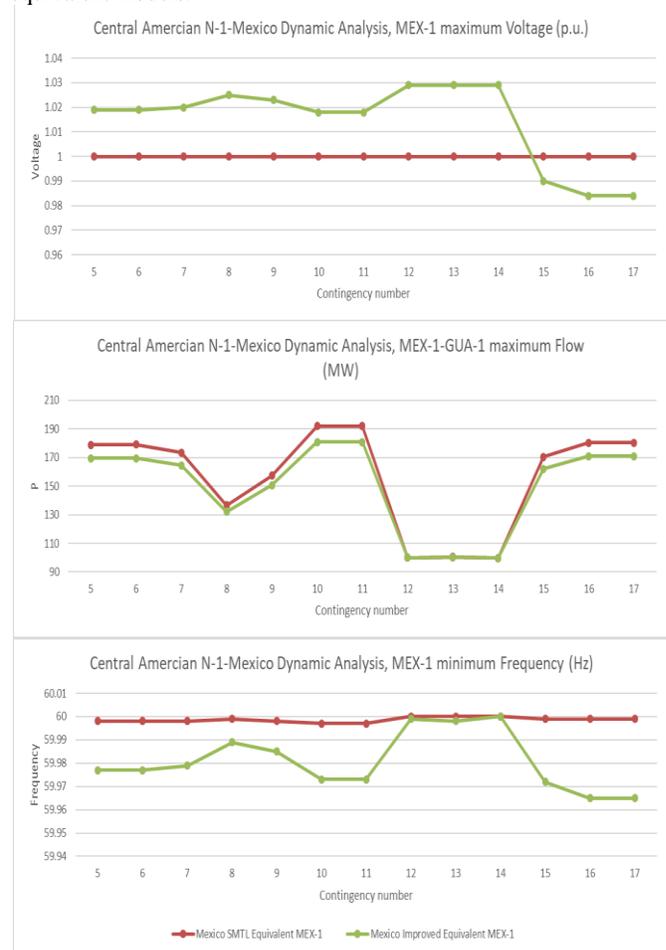

Figure 5. Result comparison for two different Mexico equivalents in Central American N-1-Mexico dynamic analysis.

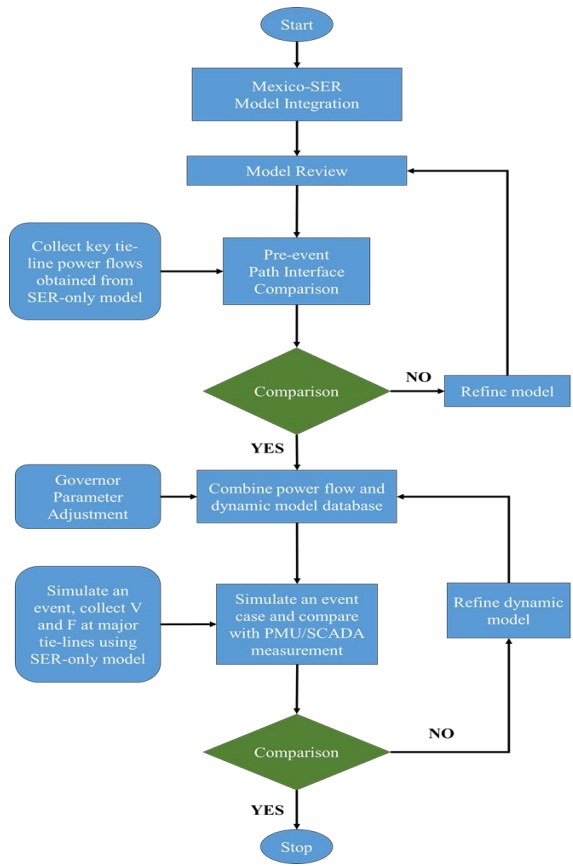

Figure 6. The proposed wide-are model validation procedure for CA+Mexico interconnected power systems.

## IV. CASE STUDY: A REAL EVENT IN 2017

A real system disturbance that occurred in March 2017 was analyzed with the proposed methodology, and field measurements were used for performance comparison.

Simulation results were analyzed in a 75-second simulation following the initial line tripping in Panama. Frequency variations were checked at GUA-2 (north block) and PAN-1 (south block) to understand representative behavior in north and south. Comparing simulation results with measurement data, following conclusions can be drawn.

1) Simulation results agree with PMU measurements in terms of shape of curves. Although numeric differences exist in maximum and minimum values, their times of occurrence are comparable.

2) Results obtained from CA+Mexico and CA-only power systems model are similar. But, the integrated model is able to better describe the frequency and voltage behavior at MEX-1. The integrated model, hence, is useful for future planning and operation studies in the region which needs an accurate representation of the electrical behavior in MEX-1.

3) The CA+Mexico integrated model is able to predict increase in power flow from Panama to Costa Rica after the Line *P1* trip. The model is also able to predict the rapid increase in power flow over the NIC-1 and CR-1 line after MEX1 – GUA1 trip.

4) Additional analysis has been performed based on unusual observations at NIC-1 and CR-1 substations, which includes the voltage and power flow comparison, as well as distance relay impedance calculation based on simulation results and PMU data.

Several areas for model refinement have been found and recommended to the regional authorities. These include tuning dynamic model parameters as well as integrating full protection relay and RAS models. Evaluating the feasibility of implementing a RAS at the Panama-Costa Rica interface is also recommended, which may act if power injection from Panama in a south-north direction increases beyond a threshold; this would arrest disturbances originating in Panama from propagating further throughout the system by preventing overload at line *N1*. Another challenge faced in this work was limited visibility of conditions in Nicaragua. The feasibility of installing at PMU at line *N1* should be explored, since it could greatly help in reconstructing precise sequence of regional events and evaluation of interarea transfers.

## V. CONCLUSION

This paper demonstrates an automated model validation procedure for Central American and Mexico interconnected systems models, and reiterates the importance of emphasizing on qualitative aspects for wide-area model validations. Several opportunities for refining the interconnected Central America and Mexico model have been identified, and recommendations for improving system stability has been put forth. The integrated model developed and validated during this study will be valuable for future studies like regional protection models and renewable energy integration. Moreover, the validation procedure developed herein will be incorporated into EOR's model validation practices, to continuously improve SER's dynamic model and therefore, its reliability.